\documentclass[reprint,aps,twocolumn]{revtex4}
\usepackage{amsmath}
\usepackage{hyperref}
\usepackage{graphicx}
\usepackage{dcolumn}
\usepackage{bm}
\ruledtabular
\begin{document}

\title{Sign reversal of magnetoresistance and p to n transition in Ni doped ZnO thin film}
\author{Arpana Agrawal$^{a}$}
\altaffiliation{Presently at Department of Physics, Indian Institute of Technology, Bombay-400076, Mumbai, Maharashtra, India}
\author{Tanveer A. Dar$^{b}$}
\email{tanveerphysics@gmail.com}
\author{R.J. Choudhary$^{c}$} \author{Archana Lakhani$^{c}$} \author {Pranay K. Sen$^{b}$}\author {Pratima Sen$^{a}$}
\affiliation{ $^{a}$School  of Physics, Devi Ahilya University, Takshashila Campus, Indore - 452001, India}
\affiliation{$^{b}$Department of Applied Physics and Optoelectronics, Shri. G. S. Institute of Technology and Science, Indore - 452003, India}
\affiliation{$^{c}$UGC-DAE Consortium for Scientific Research, Khandwa Road, Indore - 452001, India}

\begin{abstract}
We report the magnetoresistance and nonlinear Hall effect studies over a wide temperature range in pulsed laser deposited Ni$_{0.07}$Zn$_{0.93}$O thin film. Negative and positive contributions to magnetoresistance  at high and low temperatures have been successfully modeled by the localized magnetic moment and two band conduction process involving heavy and light hole subbands, respectively. 
Nonlinearity in the Hall resistance also agrees well with the two channel conduction model. A negative Hall voltage has been found for T$\geq$50\,K, implying a dominant conduction mainly by electrons whereas positive Hall voltage for T$<$50\,K shows hole dominated conduction in this material. Crossover in the sign of magnetoresistance from negative to positive reveals the spin polarization of the charge carriers and hence the applicability of Ni doped ZnO thin film for spintronic applications.

\end{abstract}
\maketitle 
Magnetoresistive random access memory, spin valves etc., utilizing magnetoresistance (MR) phenomena have been successfully fabricated.
Control of spin degree of freedom of electrons along with the charge in diluted magnetic semiconductors (DMSs) is presently an active field of research \cite{Peters}. Magnetic impurity doped ZnO based DMSs are currently the most important class of materials for investigation of spintronic applications because of the presence of small spin-orbit coupling 
 and room temperature ferromagnetism \cite{Lee,Hsu,Althammer}. 

MR measurement is a powerful tool to investigate the spin dependent functionality like spin polarized transport in DMSs where both positive and negative MR are found to exist depending on the transition metal ion doping and measuring temperatures \cite{Xu1,Kim1,Venkatesh1,Wang2}. Several interpretations have been suggested so far regarding the occurrence of negative(positive) MR at high(low) temperatures, viz; weak localization/antilocalization \cite{Xu, Choo}, 
Lorentz force \cite{Tian}, 
spin split bands formed due to Zeeman effect \cite{Khosla}
multichannel carrier conduction \cite{Kim}, etc. Despite these encouraging findings, origin of MR particularly positive MR at low temperatures in several ZnO based compounds is still controversial and lead researchers to look for the existence of nonlinearity in the Hall resistance. Nonlinearity in field dependent Hall resistivity is reported to have its origin in magnetic ordering and is identified as anomalous Hall effect (AHE) \cite{Hsu,Can}. 

MR results were also reported to be significantly affected by density, type of charge carriers and the band structure modification \cite{Xu1,Xu,Vitoratos,Michel,Chen}. Xu et al. \cite{Xu1} studied the carrier concentration dependent MR in Co doped ZnO films and found metal-to-insulator transition around the critical carrier concentration (n$_{c}$=10$^{19}$\,cm$^{-3}$) 
 \cite{Xu1}. Wu and Chen \cite{Chen} have reported the effect of energy band structures on MR of degenerate semiconductors in strong magnetic fields. Furthermore, a number of investigations have also been devoted to the study of MR in individual undoped and n/p-type doped ZnO thin films \cite{Agrawal,Andrearczyk,Herng,Wang}, although there is a lack of literature related to the study of MR in doped ZnO thin films which exhibit both n-type and p-type character at different temperatures. 
 Systems with holes as the majority charge carriers seem to be of great importance because of their large effective mass and strong spin-orbit coupling. 

In view of this, present work aims to study MR and Hall effect in Ni$_{0.07}$Zn$_{0.93}$O film and to investigate the origin of positive MR and nonlinearity in the Hall effect.

Ni$_{0.07}$Zn$_{0.93}$O thin film (thickness = 200\,nm) was grown on quartz substrate using pulsed laser deposition technique. Further details of the deposition technique are given elsewhere \cite{Agrawal}.
From the X-ray diffraction results (not shown here), the grown film was found to be c-axis oriented, single phase and preserves the wurtzite structure of ZnO. Isothermal MR measurements were performed in four probe geometry where an in-plane magnetic field was applied parallel to the direction of current, giving rise to longitudinal MR. Hall measurements in the temperature range 2-300\,K and in the field range 0-5\,T were carried out using four probe technique in 9\,T ACT PPMS, system from Quantum Design, USA.

Figure 1 represents the isothermal MR data taken upto $\pm$8\,T magnetic field (H) and clearly shows that the values as well as the character of MR are highly sensitive to the temperature (T) and magnetic field. 
At low temperature values ($<$50\,K), initially MR reduces with increase in H values and attains a minimum value at certain field beyond which it starts increasing with further increase in field and crosses to a positive value. While at higher temperatures ($\geq$50\,K), MR is always negative and its magnitude increases with increasing magnetic field. In order to understand the mechanism responsible for such behavior, MR curves were fitted taking into account both positive and negative contribution \cite{Khosla}. Symbols shown in fig. 1 are the experimental data points while the solid thick lines are the best theoretical fits given by \cite{Khosla},
\begin{figure} 
\centering
\includegraphics[width = 7cm,height=5.5cm]{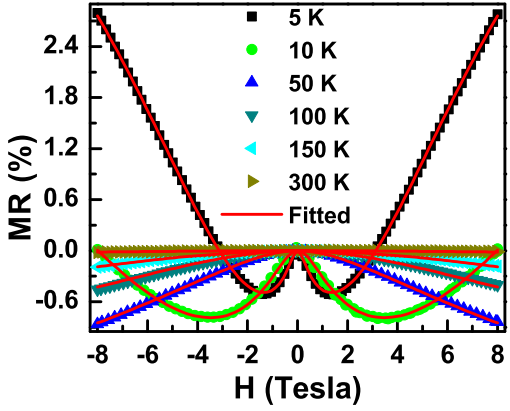}
\caption{MR data taken at various constant temperatures. Thick solid lines are the best theoretical fits to the experimental curves using equation (1).}%
\label{LMR}
\end{figure}
 \begin{equation}
\frac{\Delta\rho}{\rho_{0}}= -a^{2}\ln{(1+b^{2}H^{2})}+\frac{c^{2}H^{2}}{1+d^{2}H^{2}} 
\end{equation}  
here $\Delta\rho$ (=($\rho_{H}$-$\rho_{0}$)) is the change in resistivity due to applied magnetic field, $\rho_{H}$($\rho_{0}$) being the electrical resistivity in presence(absence) of applied field, respectively and a, b, c and d are the fitting parameters given by,
\begin{equation}
a^{2}=A_{1}J\rho_{F}\left[S(S+1)+\langle{M^{2}\rangle}\right]
\end{equation}
\begin{equation}
b^{2}=\left[1+4s^{2}\pi^{2}\left(\frac{2J\rho_{F}}{g}\right)^{4}\right]\left(\frac{g\mu_{B}}{\alpha K_{B}T}\right)^{2}
\end{equation}
\begin{equation}
c^{2}=\sigma_{1}\sigma_{2}\left(\frac{\mu_{1}+\mu_{2}}{\sigma_{1}+\sigma_{2}}\right)^{2}; d^{2}= \left(\frac{\sigma_{1}\mu_{2}-\sigma_{2}\mu_{1}}{\sigma_{1}+\sigma_{2}}\right)^{2}
\end{equation}
Here J is the exchange interaction energy, g is the Lande factor, $\rho_{F}$ is the density of states at Fermi level, $\langle{M^{2}}\rangle$ is the average magnetization of localized ions, S is the spin of localized magnetic moment, $\alpha$ is the numerical constant ranging from 0.1 to 10 and the parameter A is a measure of contribution of spin scattering to the total MR. c and d being the specific parameters related to the conductivity ($\sigma_{i}$) and mobility ($\mu_{i}$) (i=1,2) of different group of carriers in the two subbands. 
 \begin{figure} 
\centering
\includegraphics[width = 8.5cm,height=4.6cm]{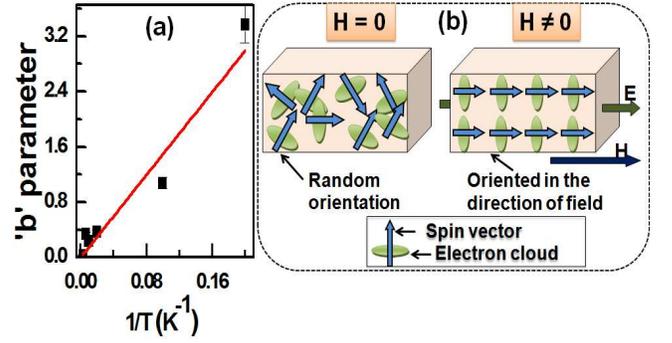}
\caption{(a) Inverse temperature dependency of the 'b' parameter; (b) Schematic depicting the localized magnetic moment model; Random orientation of the magnetic moments in the absence of external magnetic field (left) and oriented magnetic moments in the direction of applied magnetic field (right).}%
\label{bpara}
\end{figure} 

Equation 1 exactly replicate the experimental curves, suggesting an excellent agreement between the theory and experimental MR behavior at all measuring temperatures. 
To check the validity of the applied model, it may be noted that, parameter b has a 1/T dependency (fig. 2(a)).
First term in eq. (1) is dominant at high temperatures and corresponds to negative MR which takes into account the contributions of localized magnetic moments arises from magnetic impurity. In the absence of applied field, these localized magnetic moments are randomly oriented causing more scattering of the charge carriers and hence enhanced resistance (fig. 2b)(left)). When a field is applied, these randomly oriented magnetic moments get aligned in the direction of field, thereby reducing the scattering of charge carriers (fig. 2b(right)). 

We now address the behavior of positive MR at 5\,K and 10\,K, where the second term of eq. (1) becomes significant and suggests active role of charge carriers. Khosla and Fischer have attributed the cause of positive MR to the spin split bands due to Zeeman effect \cite{Khosla}. Multichannel conduction mechanism involving different electronic bands and/or spatially separated parallel conducting channels, can also give rise to positive MR \cite{Kim,Chasapis}. Such conduction mechanism has also been observed in various doped semiconductors where electron and hole subbands are considered \cite{Connell,Wang1}. Chikoidze et al. \cite{Chikoidze} have also reported positive MR in 1\% Ni doped ZnO film at 300\,K with a continous increase with field
\cite{Chikoidze}.

Figure 3(a)-(f) shows the magnetic field dependency of change in Hall resistance ($\Delta$R$_{xy}$(H)) at various temperatures. Two important features were observed, (i) negative (positive) R$_{xy}$(H) for T$\geq$50\,K (T$<$50\,K) and (ii)
occurrence of nonlinearity at low temperatures.
\begin{figure}
\centering
\includegraphics[width = 8.5cm,height=5.2cm]{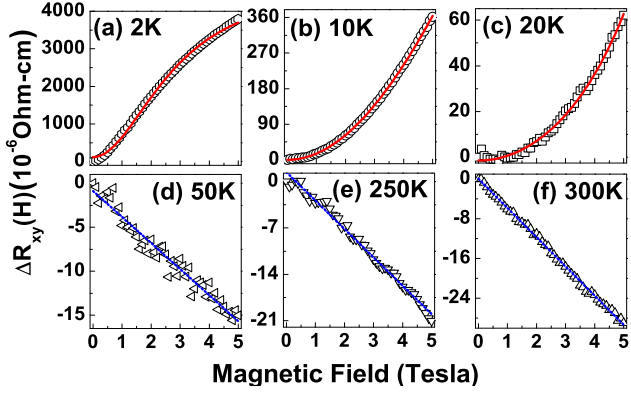}
\caption{(a)-(f) Magnetic field dependency of change in Hall resistivity of Ni$_{0.07}$Zn$_{0.93}$O film. (a)-(c) Solid thick lines (red color) are the best theoretical fits to the experimental data using eq. (5) for T$<$50\,K. (d)-(f) Dashed thick lines (blue color) are the linear fits to the experimental data for T$\geq$50\,K.}%
\label{Rh}
\end{figure}
At first glance, because of the appearance of nonlinearity in the R$_{xy}$(H) (fig. 3(a-c)), it is tempting to consider AHE, often found in ferromagnetic materials. 
Hall resistance describing the AHE consists of two terms, R$_{xy}$(H) = R$_{o}$H + R'$_{xy}$, where R$_{o}$ is the ordinary Hall coefficient and R'$_{xy}$ (=R$_{a}$$\mu$$_{o}$M) is the anomalous Hall resistance, usually, proportional to magnetization (M) with R$_{a}$ being the anomalous Hall coefficient \cite{Hsu}. If the nonlinearity results from a ferromagnetic origin, then R$_{xy}$(H) and the magnetoresistivity are expected to show the magnetic hysteresis which is not visible in the present case.

In our case, the origin of positive MR may lie in the two band model involving two parallel conducting channels. While studying nonlinear Hall effect and multichannel conduction in LaTiO$_{3}$/SrTiO$_{3}$ superlattices, Kim et al \cite{Kim} have fitted their data by the following equation,
 \begin{equation}
R_{xy}(H)= \frac{p+q^{2}H^{2}}{r+s^{2}H^{2}}.
\end{equation} 

Based on this equation, we have fitted the R$_{xy}$(H) data with the constraint of R$_{xx}$(0))=1/e($\mu_{1}n_{1}+\mu_{2}n_{2}$). Here, e is the electronic charge, n$_{i}$ (i=1, 2) is the carrier density and p, q, r, and s are the fitting parameters defining the relation between the carrier densities and the carrier mobilities in the two subbands and are defined as;
\begin{equation}
p = \mu_{1}^{2}n_{1}+\mu_{2}^{2}n_{2}; q = \frac{\mu_{1}^{2}\mu_{2}^{2}}{n_{1}+n_{2}}.
\end{equation} 
\begin{equation}
r = e(\mu_{1}n_{1}+\mu_{2}n_{2})^{2}; s = e\left(\frac{\mu_{1}^{2}\mu_{2}^{2}}{(n_{1}+n_{2})^{2}}\right).
\end{equation} 
Excellent fitting at low temperatures suggests that the two band conduction process is responsible for such behavior in R$_{xy}$(H) and hence the positive nonlinear MR.
\begin{figure} 
\centering
\includegraphics[width = 8.4cm,height=4.4cm]{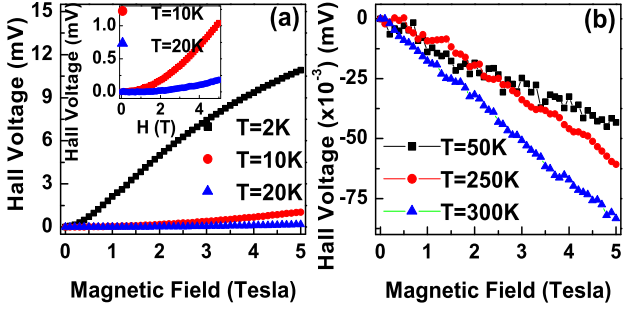}
\caption{Magnetic field response of the Hall voltage of Ni$_{0.07}$Zn$_{0.93}$O film; (a) and (b) shows the results for T$<$50\,K and T$\geq$50\,K, respectively. Inset in (a) shows the nonlinear behavior of Hall voltage with field at 10\,K and 20\,K. }%
\label{VH}
\end{figure}
\begin{figure} [htb]
\centering
\includegraphics[width = 8.4cm,height=5.6cm]{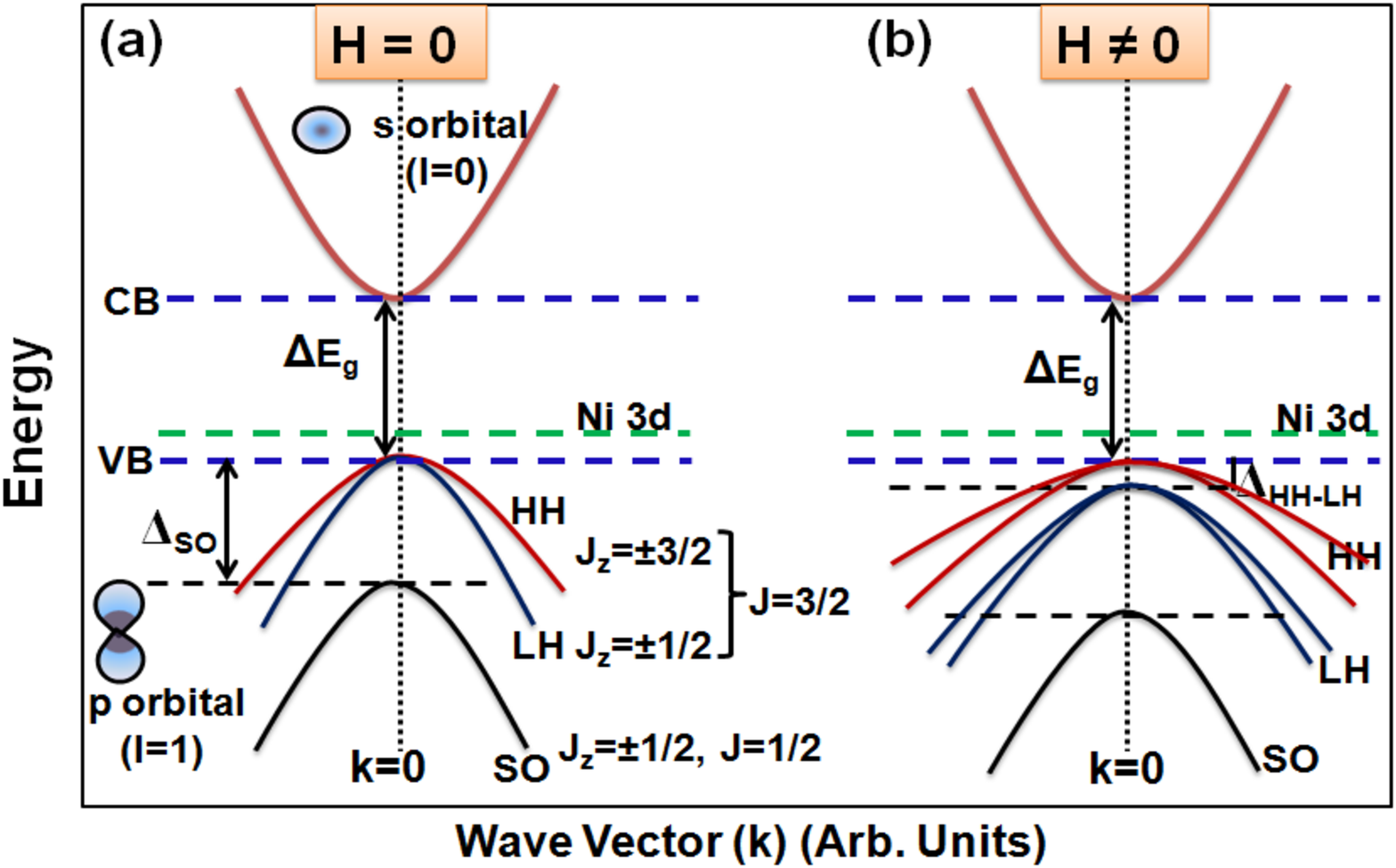}
\caption{Schematic band structure of NiZnO. (a)With conduction band (CB) (top), degenerated heavy hole (HH) and light hole (LH) bands (middle), and spin split off (SO) bands (bottom) in the absence of external magnetic field (H=0), (b) Shows the lift in degeneracy of HH and LH at k=0.}%
\end{figure}

Figure 4 demonstrates the magnetic field dependency of the Hall voltage (V$_{H}$). From fig. 4, it is evident that the slope of V$_{H}$ switches from negative to positive as the temperature goes down below 50\,K. 
Positive V$_{H}$ (thus the Hall coefficient) at low temperatures (fig. 4(a)) (T$<$50\,K) signatures the p-type conductivity of Ni$_{0.07}$Zn$_{0.93}$O film implying holes are the majority charge carriers whereas the negative sign of V$_{H}$ for T$\geq$50\,K (fig. 4(b)) reveals that electrons are the majority charge carriers and hence the n-type conductivity of the film at high temperatures. These observations show that Ni doping is akin to hole doping in ZnO as a ground state. It is worth mentioning here that both holes and electrons have different mobility and hence they have different magnetic field and voltage responses. They will also behave differently at different temperatures. Also, the nonlinearity in V$_{H}$ for T$<$50\,K is in corroboration with the low temperature MR data. 
 Electron density is estimated to be of the order of 10$^{20}$\,cm$^{-3}$ (T$\geq$50\,K), 
while the hole density is found to be in the range 10$^{17}$-10$^{19}$\,cm$^{-3}$ (T$<$50\,K). 

Keeping in view the p-type nature at low temperatures, we now concentrate on the valence band of NiZnO, formed by O 2p and Ni 3d states. The valence band further comprises of three subbands \textit{viz;} heavy hole (HH) (with J=3/2; J$_{z}$=$\pm$3/2), light hole (LH) (J=3/2; J$_{z}$=$\pm$1/2) and spin split off (SO) (J=1/2; J$_{z}$=$\pm$1/2) bands. In the absence of external magnetic field and strain, these HH and LH bands are degenerate at the center of the Brillioun zone (fig. 5a). On applying magnetic field, these subbands experience Zeeman splitting due to p-d exchange interaction with the Ni ion. Consequently, the degeneracy breaks resulting in the change in the relative energy positions of the subbands (fig. 5b). Effective masses of LH and HH differ leading to two hole species having different mobilities. 
At low temperatures, the magnetic field induced splitting of HH and LH bands is larger than the thermal energy, giving rise to spin polarized current. This feature vanishes at higher temperatures where the thermal energy dominates over the splitting energy causing disappearance of spin polarization of current. Similarly, the disappearance of polarization at higher fields at 5\,K and 10\,K arises due to the fact that energy bands corresponding to +3/2, +1/2 and (-3/2, -1/2) at positive (negative) magnetic fields come closer to the valence band and holes from both the bands contribute to the transport. Since the spin of these bands are a mixture of $\pm$ 1/2, therefore spin polarization disappears at higher fields. The crossover of the sign of MR from negative to positive also reveals the possibility of spin polarized current \cite{Peleckis}.

In conclusion, positive(negative) MR and p(n)-type conductivity have been observed at low(high) temperature regimes in Ni$_{0.07}$Zn$_{0.93}$O film. Positive MR and nonlinearity in the Hall resistance is explained on the basis of two band model comprising of LH and HH bands. Switching of MR from negative to positive at temperatures below 50\,K are also reported.

Authors are thankful to Dr. M. Gupta, Dr. R. Rawat and Mr. Sachin Kumar, UGC-DAE CSR, Indore for providing XRD and isothermal MR measurements, respectively. AA is thankful to Prof. S. Dhar, Indian Institute of Technology, Bombay for fruitful discussions. Financial supports received from MPCST, Bhopal and UGC-DAE CSR, Indore are highly acknowledged.



\begin{thebibliography}{[99]}
\bibitem{Peters}
J.A. Peters, N.D. Parashar, N. Rangaraju, and B.W. Wessels, 
Phys. Rev. B \textbf{82}, 205207 (2010).

\bibitem{Lee}
H.J. Lee, E. Helgren, and F. Hellman, 
Appl. Phys. Lett. \textbf{94}, 212106  (2009).

\bibitem{Hsu}
H.S. Hsu, C.P.Lin, H. Chou, and J.C.A. Huang, 
Appl. Phys. Lett. \textbf{93}, 142507 (2008).


\bibitem{Althammer}
M. Althammer, E.M. K-Müller, S.T.B. Goennenwein, M. Opel, and R. Gross, 
Appl. Phys. Lett. \textbf{101}, 082404 (2012).

\bibitem{Xu1}
Q. Xu, L. Hartmann, H. Schmidt, H. Hochmuth, M. Lorenz, R. Schmidt-Grund, C. Sturm, D. Spemann, and M. Grundmann, 
Phys. Rev. B \textbf{73}, 205342 (2006).

\bibitem{Kim1}
W.H. Kim, and J.Y. Son, 
Mater. Lett. \textbf{133}, 101 (2014).

\bibitem{Venkatesh1}
S. Venkatesh, A. Baras, J.S. Lee, and I.S. Roqan, 
AIP Adv. \textbf{6}, 035019 (2016).


\bibitem{Wang2}
X.L. Wang, Q. Shao, A. Zhuravlyova, M. He, Y. Yi, R. Lortz, J.N. Wang, and A. Ruotolo, 
Sci. Rep. DOI: 10.1038/srep09221.


\bibitem{Xu}
Q. Xu, L. Hartmann, H. Schmidt, H. Hochmuth, M. Lorenz, R. Schmidt-Grund, C. Sturm, D. Spemann, and M. Grundmann, 
Phys. Rev. B \textbf{76}, 134417 (2007).

\bibitem{Choo}
S.M. Choo, K.J. Lee, S.M. Park, J.B. Yoon, G.S. Park, C.Y. You, and M.H. Jung, 
Appl. Phys. Lett. \textbf{106}, 172404 (2015).


\bibitem{Tian}
Y. Tian, W. Lin, and T. Wu, 
Appl. Phys. Lett. \textbf{100}, 052408 (2012).

\bibitem{Khosla}
B.P. Khosla and J.R. Fischer, 
Phys. Rev. B \textbf{2}, 4084 (1970).


\bibitem{Kim}
J.S. Kim, S.S.A. Seo, M.F. Chisholm, R.K. Kremer, H.U. Habermeier, B. Keimer, and H.N. Lee, 
Phys. Rev. B \textbf{82}, 201407(R) (2010).

\bibitem{Can}
M.M. Can, S.I. Shah, and T. Firat, 
Appl. Surf. Sci. \textbf{303}, 76 (2014).

\bibitem{Vitoratos}
E. Vitoratos, and S. Sakkopoulos, 
J. Physique Colloques \textbf{49}, C8-193 (1988).

\bibitem{Michel}
C. Michel, P.J. Klar, S.D. Baranovskii, and P. Thomas, 
Phys. Rev. B \textbf{69}, 165211 (2004).

\bibitem{Chen}
C.C. Wu, and A. Chen, 
Phys. Rev. B \textbf{21} (1980).

\bibitem{Agrawal}
A. Agrawal, T.A. Dar, P. Sen, and D.M. Phase, 
J. Appl. Phys. \textbf{115}, 143701 (2014).

\bibitem{Andrearczyk}
T. Andrearczyk, J. Jaroszyski, and G. Grabecki, 
Phys. Rev. B \textbf{72}, 121309(R) (2005).

\bibitem{Herng}
T.S. Herng, S.P. Lau, L. Wang, B.C. Zhao, S.F. Yu, M. Tanemura, A. Akaike, and K.S. Teng, 
Appl. Phys. Lett. \textbf{95}, 012505 (2009).

\bibitem{Wang}
Y. Wang, X. Luo, L.T. Tseng, Z. Ao, T. Li, G. Xing, N. Bao, K. Suzukiis, J. Ding, S. Li, and J. Yi, 
Chem. Mater. \textbf{27}, 1285 (2015).


\bibitem{Chasapis}
T.C. Chasapis, Y. Lee, E. Hatzikraniotis, K.M. Paraskevopoulos, H. Chi, C. Uher, and M.G. Kanatzidis, 
Phys. Rev. B \textbf{91}, 085207 (2015).


\bibitem{Connell}
J.G. Connell, J. Nichols, J.H. Gruenewald, D.W. Kim, and S.S.A. Seo, 
Sci. Rep. DOI: 10.1038/srep23621.

\bibitem{Wang1}
X.R. Wang, L. Sun, Z. Huang, W.M. Lü, M. Motapothula, A. Annadi, Z.Q. Liu, S.W. Zeng, T.Venkatesan, and Ariando, 
Sci. Rep. DOI: 10.1038/srep18282.

\bibitem{Chikoidze}
E. Chikoidze, M. Boshta, M.H. Sayed, and Y. Dumont, 
J. Appl. Phys. \textbf{113}, 043713 (2013).


\bibitem{Peleckis}
G. Peleckis, X.L. Wang, and S.X. Dou, 
IEEE T Magn. \textbf{42} (2006).


\end{thebibliography}
\end{document}